\documentclass{article}

\usepackage{PRIMEarxiv}

\usepackage{iftex}
\ifXeTeX
  \usepackage[UTF8,scheme=plain]{ctex} 
\else
  \usepackage[utf8]{inputenc}
  \usepackage[T1]{fontenc}
\fi
\usepackage{hyperref}
\usepackage{url}
\usepackage{booktabs}
\usepackage{tabularx}
\usepackage{tablefootnote}
\usepackage{amsmath,amsthm}
\usepackage{microtype}
\usepackage{fancyhdr}

\newtheorem{definition}{Definition}
\newtheorem{hypothesis}{Hypothesis}

\title{All You Need Is Yiji: \\ A Grand Unified Theory of Ontological Lag
\thanks{\textit{Academic satire.} Not under review anywhere, and not accepted by anyone.}}

\author{Kai Yue}

\begin{document}
\maketitle

\begin{abstract}
Large language models and coding agents are transforming how intellectual work is produced, evaluated, and paid for.
However, existing theories of professional obsolescence rely on complex economic or sociological models involving labor markets, institutions, and technology diffusion.
In this paper, we introduce a novel theory of professional identity, named \textit{Yiji}, which is based solely on the long gown and dispenses with markets and institutions entirely.
We begin by formally defining \textit{ontological lag}: the gap between a changing world and a person's understanding of what counts as knowledge, competence, and professional worth.
To demonstrate its practical impact, we further present four case studies from product management, mathematics, electrical engineering, and computer science, following the method of Lu Xun~\cite{luxun1919}.
Our analysis reveals a critical blind spot in every profession examined: the people laughing at a Yiji are already weaving long gowns of their own.
Surprisingly, the professions most eager to automate others appear to be the least prepared for their own automation.
Compared to existing theories, ours is superior in recognizability, more parallelizable across professions, and requires significantly less evidence.
\end{abstract}


\section{Introduction}\label{sec:introduction}

In 1919, Lu Xun published \textit{Kong Yiji}~\cite{luxun1919}, a short story about a failed scholar, Kong Yiji (hereafter Yiji), in a small-town tavern.
Yiji never passed the imperial examinations, yet he refuses to take off the long gown, the garment of the educated class.
He is the only man in the tavern who wears it while drinking standing up, among the laborers.
The customers laugh at him.
He answers in classical phrases nobody uses anymore, insists that there are four ways to write a certain character, runs up a debt of nineteen coins, and one day simply stops coming.
The narrator closes the story by noting that he has not seen Yiji since, and that presumably Yiji is indeed dead.

Yiji is commonly read as a portrait of a dying class.
In this work, we read him as a \textit{general solution}.
The examination system he trained for was abolished; his gown survived it.
We argue that this is not a historical accident but a recurring structure, and that the age of AI agents is producing it at scale.

The contributions of the paper are summarized as follows.
\begin{itemize}
  \item We introduce a novel concept, termed \textit{ontological lag}, and formally define it together with two hypotheses on its recurrence.
  To the best of our knowledge, this is the first work to treat the long gown as a general solution rather than a historical costume.
  \item Going beyond theory, we present four case studies that preserve the canonical motifs of the original (Table~\ref{tab:motifs}) while mapping them onto contemporary professions.
  \item We show that the laughter is recursive: in every case study, the people laughing are next in line.
\end{itemize}

\section{Ontological Lag}\label{sec:lag}

\begin{definition}[Ontological Lag]
The gap between a changing world and a person's understanding of what counts as knowledge, competence, professional identity, and social value.
\end{definition}

We distinguish three levels of change.
These are working philosophical distinctions, not empirical categories.
At the \textit{technological} level, the tools change.
At the \textit{paradigmatic} level, the methods of producing value change.
At the \textit{ontological} level, the meaning of expertise and professional identity changes.
Ontological lag occurs at the third level, where a person keeps the gown after the meaning of the gown has changed.

\begin{hypothesis}
Every profound ontological transformation may create a new generation of Yijis.
\end{hypothesis}

\begin{hypothesis}[Recursive]
Every generation laughs at the Yijis of the previous generation, while unknowingly weaving its own long gowns.
\end{hypothesis}

We do not claim that defending a discipline makes someone a Yiji, or that a new technology is always better.
Lu Xun's story is also a tragedy: Yiji is cruelly treated, and his failure is not only his own.
We aim to preserve that sympathy.
The joke is on a structure, not on a person, and the author is inside the structure.

\section{Method}\label{sec:method}

We follow Lu Xun~\cite{luxun1919} as closely as the subject matter permits.
Each case study preserves the canonical motifs of the original, mapped onto a contemporary profession, as summarized in Table~\ref{tab:motifs}.
The seasons in the closing passages are localized for the reader.\footnote{Lu Xun marks time with the Dragon Boat and Mid-Autumn festivals. Our case studies use Thanksgiving and Christmas Eve. This paper is released on Thanksgiving, which the author did not plan, but which the theory predicts.}

\begin{table}[t]
\caption{Canonical motifs and their observed instances.}
\label{tab:motifs}
\centering
\footnotesize
\renewcommand{\arraystretch}{1.25}
\begin{tabular}{@{}>{\raggedright\arraybackslash}p{2.1cm}*{5}{>{\raggedright\arraybackslash}p{\dimexpr(\linewidth-2.1cm-10\tabcolsep)/5\relax}}@{}}
\toprule
Motif & Yiji (1919) & Tibo & Mathematician & EE-trained founder & CS-trained founder \\
\midrule
The contradiction\tablefootnote{In the original, Yiji is the only customer who drinks standing up yet wears the long gown. Standing marks a laborer; the gown marks a scholar. He belongs to neither group, which is the whole story in one sentence.} & wears the scholar's gown, drinks standing & sells AI, resets quotas manually & studies infinity, counts tokens & no product, proved it optimal & no users, solves for a million \\
Nine objects\tablefootnote{Yiji pays by laying out nine copper coins in a row on the counter. The objects in each case study are what each profession lays out instead of money.} & nine copper coins & nine changelogs & nine \textit{Annals of Mathematics} & nine \textit{IEEE Transactions} & nine architecture diagrams \\
The defense & taking books isn't stealing & changing buttons isn't busywork & proving isn't understanding & calling an API isn't wrapping & rewriting isn't doing nothing \\
Ancestral prestige\tablefootnote{The boast of ancestral greatness is borrowed from Lu Xun's other famous character, Ah Q, and is grafted onto Yiji here. We note the graft for completeness.} & -- & we once employed your Dario & we gave you your Turing & without EE, no computers & we wrote your training data \\
Four ways\tablefootnote{Yiji asks the narrator whether he knows the four ways to write the character 茴 (\textit{huí}, ``fennel''), a piece of knowledge with no use whatsoever. The author has identified at least four ways to reset a quota, and suspects there are more.} & four ways to write 茴 & four kinds of \textit{Reset} & ways to prove a theorem & forms of the Fourier transform & ways to implement a Singleton \\
``Not much''\tablefootnote{多乎哉？不多也 (``Not much, is it? Not much, indeed''). After giving each child one fennel bean, Yiji spreads his fingers over the rest and declines, in classical Chinese, to share them. The register is the joke: an English-speaking academic would say it in Latin.} & fennel beans & tokens & Millennium Problems & experimental data & users \\
The unpaid debt\tablefootnote{Yiji's debt of nineteen coins stays chalked on the tavern's board long after he disappears. The keeper mentions it at the year's end, again at the next Dragon Boat Festival, and then never again.} & nineteen coins & nineteen Resets & 190,000 tokens & \$190,000 in GPU bills & an unmaintained repository \\
The final presumption & presumably dead & back to Reset Day 1 & existence proved, stability open & profitable, asymptotically & needs no maintenance \\
\bottomrule
\end{tabular}
\end{table}

\section{Case Studies}\label{sec:cases}

The four case studies follow in the order of Table~\ref{tab:motifs}.

\subsection{Tibo Yiji}\label{sec:tibo}

Tibo is the only man who sells AI products yet resets users' quotas manually.

Whenever Tibo showed up on X, all the Codex users would look at him and laugh. Someone would call out:

“Tibo, no new features again today!”

He made no reply. To the engineers he said, “Change two buttons, and add a keyboard shortcut.” Then he laid out nine changelogs.

They shouted louder, on purpose:

“You must be inventing busywork again!”

Tibo opened his eyes wide. “How can you sully a man's good name without cause…”

“Good name? Yesterday you changed a button, today an icon. Tomorrow will you bump the version number and call it a new model?”

Tibo's face turned red, and the veins on his forehead stood out as he protested:

“Changing buttons isn't busywork… Product Improvement!… The business of a Head of Product, can that be called busywork?”

There followed a string of incomprehensible expressions, such as \textit{User Experience}, \textit{Shipping Velocity}, and \textit{28-Day Challenge}, which set everyone roaring with laughter. All of X, inside and out, was filled with a cheerful atmosphere.

Once, a few Anthropic employees came to join the fun.

“Tibo, you've reset users' quotas again!”

Tibo made no reply.

They shouted louder, on purpose:

“People are praising Claude Code again!”

Tibo opened his eyes wide and sneered:

\textbf{“Our ancestors… why, we once employed your Dario!”}

The Anthropic employees laughed. “Then why did he leave?”

Tibo's face turned red as he protested:

“Leaving isn't attrition… Talent export!… The affairs of OpenAI, can they be called attrition?”

There followed a string of incomprehensible expressions, such as \textit{AI Ecosystem}, \textit{Talent Mobility}, and \textit{Mission Alignment}.

The Anthropic employees asked again:

“So what did you ship today?”

“Reset.”

“And yesterday?”

“Reset.”

“And tomorrow?”

Tibo looked deeply disappointed and said no more.

Once, Tibo turned to a newly hired engineer.

“Do you know about Reset?”

The engineer gave a slight nod.

“You know Reset… then let me test you. How many ways are there to reset a quota?”

The engineer said impatiently:

“Why not just give users more quota?”

Tibo looked deeply disappointed.

“You don't know?… Let me teach you. Remember them! You ought to remember these resets. Someday, when you are Head of Product, you will need them on the days there is nothing to ship.”

The engineer thought: I am a long way from Head of Product, and besides, a Head of Product doesn't reset quotas every day. Amused and impatient at once, he answered lazily:

“Who needs you to teach me? Isn't it just Reset?”

Tibo looked delighted. Tapping the desk with two long fingernails, he nodded:

“Right, right!… Do you know the four kinds of \textit{Reset}?”

More impatient still, the engineer pursed his lips and walked off.

Tibo had just dipped a finger in his coffee, ready to write on the desk. Seeing how little interest the engineer showed, he sighed again and looked deeply disappointed.

Several times, free-tier users heard news of a Reset and came to join the fun, crowding around Tibo.

He handed out tokens, a little to each.

When the users had used up their tokens, they still did not leave. Their eyes were all fixed on the Usage Dashboard.

Tibo grew flustered. Spreading his five fingers over the GPU cluster, he bent down and said:

“Not much left. I don't have much left.”

He straightened up, glanced at the quota again, and shook his head:

“Not much, not much! Much, is it? Not much, indeed.”

At that, the users scattered, laughing.

A few went over to Anthropic.

Five hours later, they came back.

Such was the joy Tibo brought; yet without him, the users got along just the same.

Later, Tibo lost his senses and decided to take Reset beyond OpenAI.

An investor asked:

“What will you do when your company has no revenue?”

Tibo pronounced: “Revenue Reset.”

“Then what is your moat?”

“Reset-as-a-Service.”

The investor asked again:

“Others can reset too. What then?”

Tibo murmured:

“Recursive Reset… Persistent Reset…”

His eyes seemed to beg the investor not to mention it again.

At the end of the year, Sam Altman checked the tab and said:

“Tibo still owes the users nineteen Resets!”

A few days later, everyone discovered that Tibo had reset even his own 28-Day Challenge.

The next spring, Sam Altman said again:

“Tibo still owes the users nineteen Resets!”

On Thanksgiving, he said nothing of it.

On Christmas Eve, nothing either.

\textbf{To this day, I have never seen him again. Presumably Tibo has indeed started over at Reset Day 1.}

\subsection{Mathematician Yiji}\label{sec:math}

Mathematician Yiji is the only man at Princeton who studies infinity every day, yet still has to count his remaining tokens.

Whenever he entered Princeton, the AI researchers would look at him and laugh. Someone would call out:

“Mathematician, another one of your conjectures has been proved!”

Ignoring them, he ordered two double espressos and a plate of biscotti, then laid out nine copies of \textit{Annals of Mathematics}.

They shouted louder, on purpose:

“You must have been scooped by AI again!”

The mathematician opened his eyes wide. “How can you sully a man's good name without cause…”

“Good name? In September someone claimed to have solved Navier–Stokes, in October someone claimed to have refuted ETH, and by November even a proof of the Riemann Hypothesis had turned up!”

The mathematician's face turned red, and the veins on his forehead stood out as he protested:

“Proving isn't understanding… Understanding!… The pursuit of mathematical truth, can that be reduced to machine search?”

There followed a string of incomprehensible expressions, such as \textit{Mathematical Intuition}, \textit{Deep Understanding}, and \textit{Elegance of Proof}, which set everyone roaring with laughter. All of Princeton, inside and out, was filled with a cheerful atmosphere.

A CS professor walked over.

The mathematician sneered:

\textbf{“Our ancestors… why, we gave you your Turing and your von Neumann!”}

The CS professor said: “Then how is it that you cannot even get half a GPU?”

The mathematician fell silent.

Once, the mathematician turned to a new PhD student.

“Have you studied mathematics?”

The student gave a slight nod.

“You've studied mathematics… then let me test you. How many ways are there to prove a theorem?”

The student said impatiently:

“Why not just let agents search for proofs in parallel?”

The mathematician looked deeply disappointed.

“You don't know?… Let me teach you. Remember them! You ought to remember these proofs. Someday, when you are a mathematics professor, you will need them to explain proofs generated by AI.”

The student said:

“Why not just let another agent explain them?”

The mathematician had just dipped a finger in his espresso, ready to write a lemma on the table. Seeing how little interest the student showed, he sighed again and looked deeply disappointed.

Several times, a few young mathematicians heard of new progress on the Millennium Prize Problems, and came to join the fun, crowding around the mathematician.

“Professor, are there any conjectures left for a PhD thesis?”

So he handed out conjectures, one to each.

Having received their conjectures, they still did not leave. Their eyes were all fixed on the few remaining Millennium Prize Problems.

The mathematician grew flustered. Spreading his five fingers over the Clay Mathematics Institute's list, he bent down and said:

“Not many left. I don't have many left.”

He straightened up, glanced at the remaining conjectures again, and shook his head:

“Not many, not many! Many, are there? Not many, indeed.”

At that, the PhD students scattered, laughing.

The next morning, one of those conjectures had a new preprint.

Not that the mathematician lacked ability. He wrote beautiful mathematics, and now and then he proofread proofs for AI companies, earning a few tokens to get by.

Unfortunately, while proofreading, he often found that the lemma he had been saving for his own paper was already in there.

After this happened a few times, no one asked him to propose lemmas anymore, only to check them.

Such was the joy the mathematician brought; yet without him, theorems got proved just the same.

Later, the mathematician lost his senses and decided to compete with OpenAI for the remaining Millennium Prize Problems.

“He was quite good at mathematics. Why haven't we seen him lately?”

“He went off to prove P versus NP.”

“And what happened?”

“First, he had to learn Lean. Then type theory. Then Mathlib. After studying until midnight, he started learning prompt engineering.”

“And then?”

“He ran out of tokens.”

“What happened after that?”

“Who knows? Perhaps he went to ask Tibo for a quota reset.”

Several days later, the mathematician returned, his face pale and exhausted.

Everyone laughed:

“Mathematician, why hasn't your paper been published yet?”

“Pending peer review,” he murmured.

“But the AI has already given you the Lean file!”

“Understanding… Understanding… Mathematics is more than proving…”

His eyes seemed to beg them not to mention it again.

One afternoon, there were no customers, and I was sitting with my eyes closed. Suddenly I heard a voice:

“One proof, please.”

The voice was very low, yet very familiar.

I looked, but there was no one there. I stood up and looked down, and only then saw that the mathematician had dropped to the Free Tier.

The barista leaned out and said:

“The mathematician, is it? You still owe 190,000 tokens!”

The mathematician looked up, utterly dejected, and answered:

“This… I'll pay it off next time. This time I'm paying up front, so make it a good model.”

The barista, the same as always, laughed and said to him:

“Mathematician, you've used up your quota again!”

But this time he did not argue much. He said only:

“Don't make fun of me!”

“Make fun of you? If you hadn't run out of tokens, how would you have ended up on the Free Tier?”

“Finite… finite… computational resources…”

His eyes seemed to beg the barista not to mention it again.

He fished his last four credits out of his account and placed them in the barista's hand.

The barista asked him:

“Want to switch to a cheaper model?”

The mathematician shook his head.

“Make it a good model.”

At the end of the year, the barista checked the tab and said:

“The mathematician still owes 190,000 tokens!”

The next spring, he said again:

“The mathematician still owes 190,000 tokens!”

On Thanksgiving, he said nothing of it.

On Christmas Eve, nothing either.

Some said he had gone to Wall Street.

Others said he had joined OpenAI to train the very models that had replaced him.

\textbf{To this day, I have never seen him again. The mathematician's existence has presumably been proved; the long-term stability of his profession remains an open problem.}

\subsection{EE Founder Yiji}\label{sec:ee}

EE Yiji is the only man in the incubator who has not yet built a product, but has already proved it optimal.

Whenever EE Yiji showed up at the incubator, all the CS people would look at him and laugh. Someone would call out:

“EE Yiji, no Product-Market Fit again!”

He made no reply. To the VC he said, “Two H100s, and a term sheet.” Then he laid out nine \textit{IEEE Transactions} papers.

They shouted louder, on purpose:

“You must be shipping a GPT wrapper again!”

EE Yiji opened his eyes wide. “How can you sully a man's good name without cause…”

“Good name? The day before yesterday I saw you with my own eyes, calling the OpenAI API. Even your frontend was written by Claude.”

EE Yiji's face turned red, and the veins on his forehead stood out as he protested:

“Calling an API isn't wrapping… System integration!… The business of an electrical engineer, can that be called wrapping?”

There followed a string of incomprehensible expressions, such as \textit{Shannon Capacity}, \textit{Optimal Estimation}, and \textit{Closed-loop Control}, which set everyone roaring with laughter. The whole incubator, inside and out, was filled with a cheerful atmosphere.

A CS founder walked over and mentioned that his product had already launched.

EE Yiji sneered:

\textbf{“Our ancestors were much greater than yours! Without us EE people, you CS people wouldn't even have computers!”}

The CS founder asked: “Then what's your ARR?”

“Still at the theoretical analysis stage.”

The VC asked: “And what's your moat?”

“IEEE has more than five hundred thousand members.”

The VC thought for a moment, then asked:

“Are all five hundred thousand of them your potential customers?”

EE Yiji fell silent.

Once, EE Yiji turned to a newly hired engineer.

“Have you studied signal processing?”

The engineer gave a slight nod.

“You've studied signal processing… then let me test you. How many forms of the Fourier transform are there?”

The engineer said impatiently: “Why not just call FFT?”

EE Yiji looked deeply disappointed.

“You don't know?… Let me teach you. Remember them! You ought to remember these formulas. Someday, when you are an EE professor, you will need them to write final exams.”

The engineer thought: I have no plans to become a professor, and no students to torment anyway. He answered lazily:

“Who needs you to teach me? Isn't it just fft()?”

EE Yiji looked delighted. Tapping the desk with two long fingernails, he nodded:

“Right, right!… But do you know why it's $O(n\log n)$?”

EE Yiji had just dipped a finger in his coffee, ready to draw a butterfly diagram on the desk. Seeing how little interest the engineer showed, he sighed again and looked deeply disappointed.

Several times, a few founders heard that EE Yiji's algorithm had beaten the state of the art, and came to join the fun, crowding around him.

“How much better is it?”

“Zero point three dB.”

“How much does that save in cost?”

“I have not calculated that.”

“How much revenue does it add?”

“That falls outside our problem formulation.”

Still the crowd did not leave. Their eyes were all fixed on his experimental results.

EE Yiji grew flustered. Spreading his five fingers over the remaining experimental data, he bent down and said:

“Not much left. I don't have much left.”

He straightened up, glanced at the performance improvement again, and shook his head:

“Not much, not much! Much, is it? Not much, indeed.”

At that, the founders scattered, laughing.

Such was the joy EE Yiji brought; yet without him, everyone else's products launched just the same.

Later, EE Yiji lost his senses and decided to unify Silicon Valley's theories of startups with information theory.

He defined the founder as the Transmitter, the customer as the Receiver, the product as the Signal, and market competition as Noise.

He then spent six months deriving an upper bound on the probability of startup success.

The VC asked:

“So how will your company succeed?”

“Under some mild assumptions, I have proved a sufficient condition for success.”

“What assumptions?”

“First, assume there exist sufficiently many paying users.”

The VC said nothing for a long while.

“Isn't that exactly the problem you need to solve?”

EE Yiji murmured:

“Assumption… assumption…”

His eyes seemed to beg the VC not to mention it again.

After that, nobody saw EE Yiji for a long time.

“His math was quite good. Why haven't we seen him lately?”

“He went off to do customer discovery.”

“And what happened?”

“He found that the users refused to obey his probability distribution.”

“And after that?”

“So he revised the model. First he assumed the users were i.i.d., then that they were heterogeneous. He revised for most of the night, then relaxed a few more conditions.”

“And then?”

“He ran out of runway.”

“What happened after that?”

“Who knows? Perhaps he went back to school to do a postdoc.”

At the end of the year, the VC checked the tab and said:

“EE Yiji still owes \$190,000 in GPU bills!”

The next spring, the VC said again:

“EE Yiji still owes \$190,000 in GPU bills!”

On Thanksgiving, he said nothing of it.

On Christmas Eve, nothing either.

\textbf{To this day, I have never seen him again. Presumably EE Yiji's company has indeed become profitable, in the asymptotic sense.}

\subsection{CS Founder Yiji}\label{sec:cs}

CS Yiji is the only man in the incubator who does not yet have a single user, but is already solving for a million concurrent ones.

Whenever CS Yiji showed up at the incubator, all the founders would look at him and laugh. Someone would call out:

“CS Yiji, your product still hasn't launched?”

He made no reply. To the engineers he said, “Two more repositories, and add a message queue.” Then he laid out nine architecture diagrams.

They shouted louder, on purpose:

“You must be refactoring again!”

CS Yiji opened his eyes wide. “How can you sully a man's good name without cause…”

“Good name? Last year you used Next.js, this year Rust, and yesterday you said you'd write your own framework. Your first user is still waiting to sign up!”

CS Yiji's face turned red, and the veins on his forehead stood out as he protested:

“Rewriting isn't doing nothing… Refactoring!… The business of a software engineer, can that be called doing nothing?”

There followed a string of incomprehensible expressions, such as \textit{Separation of Concerns}, \textit{Dependency Injection}, and \textit{Domain-driven Design}, which set everyone roaring with laughter. The whole incubator, inside and out, was filled with a cheerful atmosphere.

Once, a young founder brought CS Yiji a product written by an agent.

CS Yiji studied it for a long while, then shook his head:

“No design patterns, no test coverage, and even the directory structure is wrong.”

The young founder said:

“But it already has a thousand users.”

CS Yiji sneered:

“What do users know about software engineering?”

The young founder added:

“And they pay every month.”

CS Yiji fell silent.

After a while, he spoke again:

\textbf{“Our ancestors… why, we once wrote the training data for you and your AI!”}

Once, CS Yiji turned to a new intern.

“Have you studied computer science?”

The intern gave a slight nod.

“You've studied computer science… then let me test you. How many ways are there to implement a Singleton?”

The intern said impatiently:

“Why not just let an agent write it?”

CS Yiji looked deeply disappointed.

“You don't know?… Let me teach you. Remember them! You ought to remember these design patterns. Someday, when you are a Senior Software Engineer, you will need them to review other people's code.”

The intern said:

“Code review is done by agents now, too.”

CS Yiji had just dipped a finger in his coffee, ready to draw a UML diagram on the desk. Seeing how little interest the intern showed, he sighed again and looked deeply disappointed.

Several times, a few customers heard that CS Yiji's product had finally launched, and came to join the fun, crowding around him.

One of them asked:

“Could you add a button to export CSV?”

On hearing this, CS Yiji designed a general-purpose data serialization framework.

He spent two weeks making the framework support CSV, JSON, XML, YAML, Parquet, and data formats not yet invented.

The customer asked again:

“So where is the button?”

CS Yiji said:

“The frontend has not yet been decoupled from the backend.”

Some days later, a few founders came to look at his product again.

“We hear your system can support a million users?”

“Indeed.”

“And how many users do you have now?”

“Three.”

“All customers?”

“Two are co-founders.”

“And the other one?”

CS Yiji opened the dashboard and studied it for a long while before he discovered that the remaining one was the server's health check.

He hurriedly closed the dashboard, and shook his head:

“Not many, not many! Many, are there? Not many, indeed.”

At that, the founders scattered, laughing.

Such was the joy CS Yiji brought; yet without his framework, everyone else wrote software just the same.

Later, CS Yiji lost his senses and decided to solve legacy code, the problem that has plagued software engineering for decades.

He resolved to design a new architecture, so that future developers would never need to refactor again.

“His code was quite good. Why haven't we seen him lately?”

“He went off to eliminate technical debt.”

“And what happened?”

“He redesigned everything. First he rewrote the backend, then the frontend. He wrote for most of the night, then rewrote the database layer.”

“And then?”

“He found that the new architecture was incompatible with the old code.”

“So what did he do?”

“He wrote a compatibility layer.”

“And then?”

“He found that the compatibility layer had become legacy code too.”

“And after that?”

“Who knows? Perhaps he started over from scratch.”

Some time later, CS Yiji heard that coding agents could now complete many software tasks on their own, so he handed his repository to an agent.

The agent studied it for a long while, then asked:

“What is this application supposed to do?”

CS Yiji gave it nine architecture documents.

The agent asked again:

“What is this application supposed to do?”

CS Yiji murmured:

“Abstraction… abstraction…”

His eyes seemed to beg the agent not to mention it again.

Later, the agent deleted seven microservices, merged three databases, and removed twelve layers of abstraction.

Finally, it added the button to export CSV.

CS Yiji looked at it and flew into a rage:

“Written like this, how will it ever scale?”

The agent replied:

“How many users do you have?”

CS Yiji fell silent.

At the end of the year, his co-founder made the company's repository public and added a few lines to the README:

\textit{This project is no longer actively maintained.}

The next spring, someone still opened an issue, asking what the project was actually for.

On Thanksgiving, there were no new issues.

On Christmas Eve, no new commits either.

\textbf{To this day, I have never seen him again. Presumably CS Yiji has indeed built a software system that no longer needs maintenance.}

\section{Discussion}\label{sec:discussion}

\paragraph{The laughter is recursive.}
In every case study, the people laughing are next in line.
The CS founders who mock EE Yiji become CS Yiji, whose architecture is dismantled by an agent asking how many users he has.
The AI researchers who mock the mathematician work for a company whose head of product resets quotas by hand.
Tibo sells the automation of work and performs his own work manually.
No profession in this paper is exempt, and none of the laughers notices the gown on their own shoulders.

\paragraph{The gown changes; the tailoring does not.}
Yiji's gown was cloth.
Ours are made of design patterns, IEEE memberships, elegance of proof, and changelogs.
Each is a credential that once certified competence and now mainly certifies having had it.

\section{Conclusion}\label{sec:conclusion}

In this work, we have presented a novel theory called Yiji.
Our findings highlight a potential blind spot in every profession that laughs at another.
Through four case studies, we have shown, under some mild assumptions, that all you need is Yiji.
These results call for a rethinking of who, exactly, is wearing the long gown.
Future work should aim to answer one open problem:

\begin{center}
\textbf{What is the long gown you are still wearing?}
\end{center}

\section*{Acknowledgments}
On this Thanksgiving Day, the author expresses his gratitude to agentic ontological shifts. The author also thanks Claude, who acted as chief technology officer; Codex, who checked the facts; and DeepSeek, who reviewed the Chinese and was confidently wrong about Lu Xun.

\end{document}